\documentclass[journal]{IEEEtran}

\usepackage[cmex10]{amsmath}
\usepackage{amssymb}
\usepackage{cite}
\usepackage{graphicx}
\graphicspath{{./fig/}}
\usepackage{algorithmic}
\usepackage{setspace}
\usepackage{array}
\usepackage{fixltx2e}
\usepackage{color}
\usepackage{times,color}
\usepackage{epstopdf}

\begin{document}
\title{Increasing Flash Memory Lifetime by Dynamic Voltage Allocation for Constant Mutual Information }
\author{
Tsung-Yi Chen, Adam~R.~Williamson and Richard~D.~Wesel
\\
tsungyi.chen@northwestern.edu, adamroyce@ucla.edu, wesel@ee.ucla.edu
}

\maketitle

\begin{abstract}
\boldmath 
The read channel in Flash memory systems degrades over time because the Fowler-Nordheim tunneling used to apply charge to the floating gate eventually compromises the integrity  of the cell because of tunnel oxide degradation.  While degradation is commonly measured in the number of program/erase cycles experienced by a cell, the degradation is proportional to the number of electrons forced into the floating gate and later released by the erasing process.  By managing the amount of charge written to the floating gate to maintain a constant read-channel mutual information, Flash lifetime can be extended. This paper proposes an overall system approach based on information theory to extend the lifetime of a flash memory device. Using the instantaneous storage capacity of a noisy flash memory channel, our approach allocates the read voltage of flash cell dynamically as it wears out gradually over time. A practical estimation of the instantaneous capacity is also proposed based on soft information via multiple reads of the memory cells. 
\end{abstract}

\section{Introduction}
\label{sec:Intro}
Flash memory is ubiquitous on our keychain, in our super-thin laptop, and in the racks of enterprise storage data centers.  Unfortunately, Flash memory reliability degrades over time as a function of the amount of charge that is written into and subsequently erased from the memory cell.  This degradation (called ``wear-out'') can be understood as a time-varying noise whose variance increases with the number of electrons forced into and out of the floating gate by Fowler-Nordheim tunneling.  Wear-out becomes worse as the storage density (bits per memory cell) is increased by using denser constellations to store more information.  The reliability and lifetime problems associated with these new, higher-density Flash memories have driven research into the use of LDPC codes to improve performance.

\begin{figure}
\includegraphics[width=0.5\textwidth]{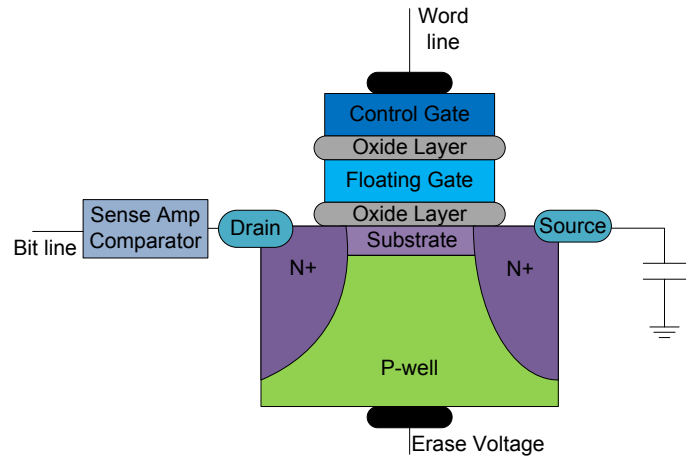}
\caption {NAND Flash memory cell structure.} \label{fig:flashcell}
\end{figure}

Fig.~\ref{fig:flashcell} illustrates the device structure of a NAND flash memory cell (i.e., a floating-gate transistor). To store information, a charge level is written to the cell by adding a specified amount of charge to the floating gate through Fowler-Nordheim tunneling by applying a relatively large voltage to the control gate~\cite{BezIEEE03}.  Actually, charge is written to the floating gate incrementally with feedback, carefully approaching the desired level from below.

To read a memory cell, the charge level written to the floating gate is detected by applying a specified word-line voltage to the control gate and comparing the transistor drain current to a threshold by a sense amp comparator. If the drain current is above the comparator threshold, then the word-line voltage was sufficient to turn on the transistor, indicating that the charge written to the floating gate was insufficient to prevent the transistor from turning off. If the drain current is below the threshold, the charge written to the floating gate was sufficient to prevent the applied word-line voltage from turning on the transistor. The sense amp comparator only provides one bit of information about the charge level present in the floating gate.

The word-line voltage required to turn on a particular transistor is called the threshold voltage. We refer to the variation of threshold voltage from its intended value as the \emph{read channel noise}. The threshold voltage can vary from its intended value for a variety of reasons. For example, the floating gate can be overcharged during the write operation, the floating gate can lose charge due to leakage in the retention period, or the floating gate can receive extra charge when nearby cells are written~\cite{MaedaISDFT09}.

This paper presents an approach that dynamically scales the amount of charge for each level  (i.e. the word-line voltage thresholds for each level) to maintain a constant mutual information over the lifetime of the device.  The motivation is to increase lifetime by  carefully managing the  precious resource of total charge written to the cell.  Our approach utilizes read-channel state information that could be provided by variable-precision decoding. Our approach maintains the simplicity of a constant instantaneous storage capacity of the device over its useful lifetime.

The rest of the paper is organized as follows: Sec.~\ref{sec:flashmodel} introduces the Flash memory channel model used in this paper. Sec. \ref{sec:voltage} presents the main ideas of our dynamic voltage allocation and the numerical results. Sec.~\ref{sec:assess} discusses practical techniques to estimate the distribution of the actual Flash memory cell using multiple reads. Finally Sec.~\ref{sec:conclusion} concludes the paper.

\begin{figure}
\centering
\includegraphics[width=0.4\textwidth]{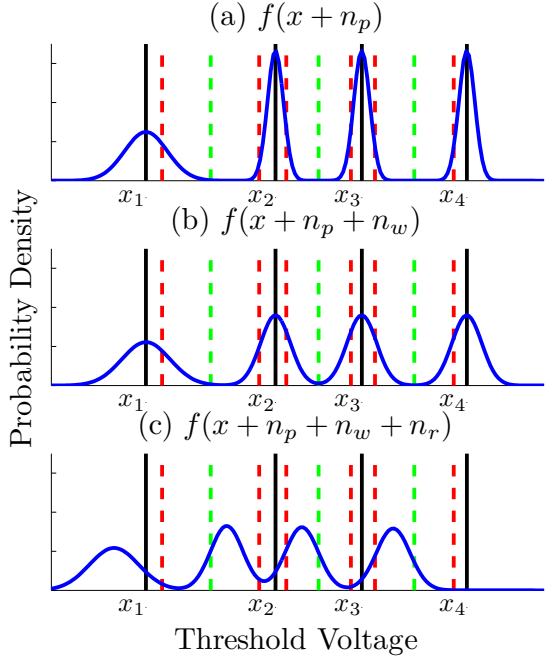}
\caption {Flash read channel model PDFs . }
\label{fig:Noise_PDFs}
\end{figure}
\section{Flash Memory Channel Model}
\label{sec:flashmodel} 
This section introduces a tractable but also realistic model for Flash memory channel. Following the general approach of \cite{Elarief_LMASE_TransIT_2010,  Klove_LMAE_TransIT_2011, Klove_LME_Flash_TransIT_2011, Elarief_LME_Zq_TransComp_2012, Cassuto_ALME_TransIT_2010} with respect to the case of limited magnitude asymmetric errors, we first design a reasonable noise model and then seek the best possible system for that model. We draw heavily upon the extensive prior theoretical investigations and experimental studies in the semiconductor device research community~\cite{Olivo-oxide-charge-86, wellekens1995write, R14_KTakeuchi_JSSC, mielke2004flash, lee2004effects, fukuda2007random, compagnoni2008first, ghetti-scaling-08, spinelli2008investigation} to capture the critical components of the time-varying and input-dependent characteristics of NAND Flash memory cells.  
Our model makes explicit the dependence of the wear-out and retention noise on the total amount of charge written to and subsequently erased from the memory cell.  We model the NAND Flash memory cell data storage process as
\begin{equation}
\label{eq:totalmodel}
y = x + n_p + n_w + n_r,
\end{equation}
where $x$ is the threshold voltage level intended to be written to the cell and $y$ is the threshold voltage when the cell is read.  The three noise components in our model are the programming noise $n_p$, the wear-out noise $n_w$, and the retention noise $n_r$, each of which is described in detail below.  Wear-out and retention noise depend explicitly on accumulated charge.

\setcounter{equation}{3}
\begin{figure*}
\begin{align}
\label{eqn:distribution}
f_{Y|X}(y|x) = \frac{e^{\sigma^2/2\lambda^2}}{2\lambda}
\left[
\exp\left(\frac{y - \mu}{\lambda}\right)Q\left(\frac{y - \mu}{\sigma} + \frac{\sigma}{\lambda}\right)
+\exp\left(\frac{-y + \mu}{\lambda}\right)Q\left(\frac{-y + \mu}{\sigma} + \frac{\sigma}{\lambda}\right)
\right].
\end{align}
\hrule
\end{figure*}
\setcounter{equation}{1}

Fig.~\ref{fig:Noise_PDFs} shows the contributions from each type of
noise. We do not include cell-to-cell interference through parasitic capacitive
coupling~\cite{R12_JDL_EDL} in our model because cell-to-cell interference is a
data-dependent process that can be partially mitigated by equalization and pre-distortion \cite{patent-yli-comp, Dong_equ}.  However, the approach can be extended to handle  an additional noise term reflecting residual cell-to-cell interference.

\subsection{Programming noise}

The programming noise $n_p$ represents memory cell threshold voltage variation immediately after a brand new cell has been written.  The programming noise is approximately Gaussian, but the variance is input-dependent~\cite{R14_KTakeuchi_JSSC, compagnoni2008first}.  Letting $x_i$ be the $i$th voltage level of a Flash cell, our model is given as
\begin{equation}
f_{N_p}(n_p)= 
\begin{cases}
\mathcal{N}(0, \sigma^{2}_{p})& \text{if } x = x_i, i > 1,\\
\mathcal{N}(0, \sigma^{2}_{e})& \text{if } x = x_1,
\end{cases}
\end{equation}
where $\sigma_{e}>\sigma_{p}$. In other words, the erased state $x_1$ has a larger noise variance.

\subsection{Wear-out noise}
Flash memory program/erase (P/E) cycling causes damage (i.e., wear-out) to the tunnel oxide of Flash memory cells in the form
of charge trapping in the oxide and interface states~\cite{Olivo-oxide-charge-86,
Cappelletti-failure-machanism-IEDM-94, mielke2004flash,yamada1993degradation}. The memory
cell wear-out caused by P/E cycling is proportional to the number of electrons tunneling through
the gate oxide that is further proportional to the programmed threshold voltage level.   Memory cell damage caused by P/E cycling is a function of the accumulated programmed threshold voltages over these P/E cycles~\cite{yamada1993degradation}. Let $V_e$ denote the voltage of the
erased state, $V_p^{(j)}$ denote the voltage of programmed state during the $j$-th P/E cycle, and
$N$ denote the total number of P/E cycles.  Define the voltage accumulated over $N$ writes as
\begin{equation}
V_\text{acc}=\sum_{j=1}^{N}(V_p^{(j)}-V_e)) \, .
\end{equation}

Based upon the discussion and measurement results presented in~\cite{fukuda2007random,
compagnoni2009random}, we model the wear-out noise $n_w$ in (\ref{eq:totalmodel})
as a Laplace $(0, \lambda)$ distribution with density $f(n_w)=\frac{1}{2\lambda}
e^{-\frac{|n_w|}{\lambda}}$ with
$\lambda = C_w+A_w \cdot \left({V_\text{acc}}/{V_\text{max}}\right)^{k_1}$
where constants $C_w$, $A_w$ and $k_1$ are technology dependent with typical values around
$1.26\times 10^{-3}$, $1.80\times 10^{-4}$ and $0.62$, respectively.

\subsection{Retention noise}

Retention noise $n_r$  models the degradation of the threshold voltage
integrity due to charge leakage after it is written.  Interface trap recovery and electron
detrapping~\cite{Mielke-recovery-effect-06,Yang-reliability-issues-06, leej2003data} gradually
reduce memory cell threshold voltages.  The degradation becomes worse as $V_\text{acc}$ becomes
larger.  Based upon the discussion and measurement results in~\cite{mielke2004flash,
Yang-reliability-issues-06}, the noise $n_r$ in (\ref{eq:totalmodel}) approximately follows a
Gaussian distribution $\mathcal{N}(\mu_r, \sigma_r^2)$, where the parameters $\mu_r$ and $\sigma_r$
are both time-varying and voltage-dependent.  Our model has mean $\mu_r$ given as
\begin{align*}
\mu_r &=-x \ln\left(1+\frac{t}{t_0}\right)\left[ A_r
\left(\frac{V_\text{acc}}{V_\text{max}}\right)^{k_1} +B_r
\left(\frac{V_\text{acc}}{V_\text{max}}\right)^{k_2}\right], 
\end{align*}
and the variance $\sigma_r^2$ given as
\begin{align*}
\sigma_r^2 &=0.1 x \ln\left(1+\frac{t}{t_0}\right)\left[ A_r
\left(\frac{V_\text{acc}}{V_\text{max}}\right)^{k_1} +B_r
\left(\frac{V_\text{acc}}{V_\text{max}}\right)^{k_2}\right]^2,
\end{align*}
where $x$ is the target threshold voltage being programmed into the memory cell.  The exponents $k_1$ and $k_1$ depend on the manufacturing technology of the devices. In the current paper we choose $k_1 = 0.62$ and $k_2 = 0.3$. The constants $A_r$ and $B_r$ are technology dependent with typical values around $7.0\times 10^{-4}$ and $4.76\times 10^{-3}$, respectively, and $t_0$ is the time constant that also depends on the manufacturing technology.

\section{Dynamic Voltage Allocation Based on Mutual Information}
\label{sec:voltage}
As discussed in Sec.~\ref{sec:flashmodel}, the noise of a Flash cell can be modeled as a sum of Gaussian and Laplacian random variables with time-varying parameters. A similar channel model has been studied in the context of multiple-user communication with impulse radio \cite{Win_TCOM_2000, Beaulieu_TCOM_2010}.  Our model channel has the overall conditional distribution given as \eqref{eqn:distribution}, shown at the top of the page, where $\sigma^2 = \sigma_p^2 + \sigma_r^2$, $\mu = \mu_r + x$ and $Q(x)$ is the tail of the standard Gaussian random variable: 
$$
Q(x) = \int_{x}^{\infty}\sqrt{2\pi}\exp\{-x^2/2\}.
$$ 
With a uniform input distribution for an $L$-level Flash cell ($L$ equally likely inputs), the distribution of the channel output is given as 
$$
f_Y(y)=\sum_{i = 1}^L f_{Y|X}(y|x_i)/L, 
$$
where $x_i$ is the $i$th voltage level and $x_1$ represents the erased state. Recall that the programming noise variance is larger than all other states as discussed in Sec.~\ref{sec:flashmodel}.

For a given voltage level, we can calculate the relevant mutual information (the highest theoretical information rate for equally likely inputs) according to the expected accumulated voltage $\mathbb{E}[V_{\text{acc}}] = \mathbb{E}[\sum_{j = 1}^N V_p^{(j)}-V_e]$ and a specified maximum retention time.   This is the \textit{instantaneous storage capacity} of a Flash cell and the total storage capacity is this value scaled by the number of cells. 

In practice, a margin between the actual channel coding rate and capacity is necessary to provide a reliable error protection. The smallest possible margin for a specified block length can be found using the finite-blocklength analysis in \cite{Polyanskiy_10}.  Specifically, for a given reliability of $\epsilon$ and the channel code's blocklength $n$, we can find the least margin required to achieve by using the finite-blocklength converse. The normal approximation formula in \cite{Polyanskiy_10} is given as
\begin{align}
nR = nC - \sqrt{n{V}}Q^{-1}(\epsilon) +O(\log n)
\end{align}
where $C$ is the channel capacity and $V$ is the channel dispersion . 
This approximation can serve as a coarse assessment of the margin between the channel coding rate and the capacity of a general Flash memory channel.

For practical high-rate LDPC codes on a Gaussian channel model, we observe experimentally that typical margins are around $0.1$ to $0.15$ bits.  Once the average value of $E[V_{\text{acc}}]$ over a group of cells is large enough that the instantaneous capacity is less than the code rate plus the needed margin, the code will fail to provide reliable error protections over that group of cells. 

There is a tension between the instantaneous capacity and the lifetime of a cell; higher voltage levels provide  higher instantaneous capacity but causes the Flash cells to wear out faster.  One possible approach is to use constant voltage levels and variable rate coding to decrease the rate as the instantaneous capacity decreases.  However, both USB drives and enterprise storage applications tend to assume that the instantaneous storage capacity remains constant over the lifetime of the device.

Our approach recognizes that the device needs to maintain a specified instantaneous storage capacity and designs the system to maximize the lifetime over which that capacity is maintained.  This does not preclude the use of variable-rate coding to allow the device to be of use during its ``after life'', but the goal is to delay this as long as possible.  We use information theory to dynamically tune the voltage levels to maintain only the needed margin, minimizing $V_{\text{acc}}$.  This achieves a longer average lifetime. The extremely large margin common in the beginning of a cell's life is not needed (since the LDPC code is in place) and that excess margin comes at the cost of a shorter lifetime. 

\begin{figure}
\includegraphics[width=0.5\textwidth]{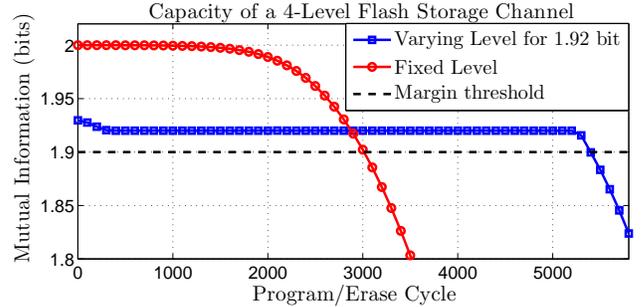}
\caption {Capacity vs. P/E cycle at $1$ year retention time.}
\label{fig:cap_vs_pec}
\end{figure}

To illustrate this, take a Flash chip with $4$-level cells as in Fig. \ref{fig:Noise_PDFs} and a rate-$8/9$ error correction code as an example. Suppose the needed margin of $0.12$ bits  requires the instantaneous capacity to be above $1.9$ bits. For this example, the technology-dependent parameters are chosen as follows: $A_w = 1.8\times 10^{-4}$, $C_w = 1.26\times 10^{-3}$, $A_r = 7.0\times 10^{-4}$, $B_r = 4.76\times 10^{-3}$, $V_{\text{max}} = 16$. The voltage levels are set to be $\{2.8, 5.2, 6.4, 7.86\}$. The time constant is set to $t_0 = 1$ (in units of hours). The state-dependent Gaussian noise variances are chosen to be $\sigma_p = 0.05$ and $\sigma_e = 0.35$. 

\begin{figure}
\includegraphics[width=0.5\textwidth]{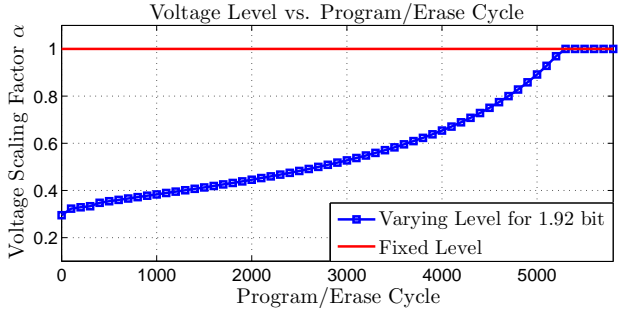}
\caption{$\alpha$ vs. P/E cycle at $1$ year retention time.}
\label{fig:VoltLevel}
\end{figure}
For a retention time of $1$ year ($t = 8760$ hours), the red curve in Fig.~\ref{fig:cap_vs_pec} is the baseline for fixed voltage levels over the lifetime. The baseline voltage levels are chosen such that the instantaneous capacity drops to $1.9$ bits at a typical lifetime of $3000$ P/E cycles for a $4$-level Flash cell. Fig.~\ref{fig:cap_vs_pec} also shows the instantaneous capacity using our dynamic voltage level approach.  The voltage levels are scaled by a single parameter $\alpha$, adjusted after every $100$ P/E cycles to attain instantaneous capacity of $1.92$ bits (slightly higher than the threshold to provide extra margin) until the voltage levels become the same as our baseline ($\alpha = 1$). This single parameter scheme improves the lifetime from $3000$ to $5400$ P/E cycles: an $80\%$ improvement in this example.  

Fig.~\ref{fig:VoltLevel} compares $\alpha$ as a function of P/E cycles between the varying level scheme and the baseline. The initial $\alpha$ is set to $0.28$ and as the device wears out, $\alpha$ is gradually increased to match the desired mutual information margin every $100$ P/E cycles. As illustrated in Fig.~\ref{fig:VoltLevel}, the lifetime improvement is significant since a brand new Flash cell only needs $28\%$ of the fixed voltage levels to achieve sufficient amount of instantaneous capacity.  Fixed voltage levels waste $70\%$ of the early voltage, needlessly damaging the Flash cell.  We emphasize that this is only an initial illustration.  Investigation of a variety of ways to improve performance further including optimizing the frequency of charge level adjustment and more carefully optimizing each of the charge levels is ongoing work.

\section{Assessment of Wear-Out and Retention Noise }
\label{sec:assess}
\begin{figure}
\includegraphics[width=0.45\textwidth]{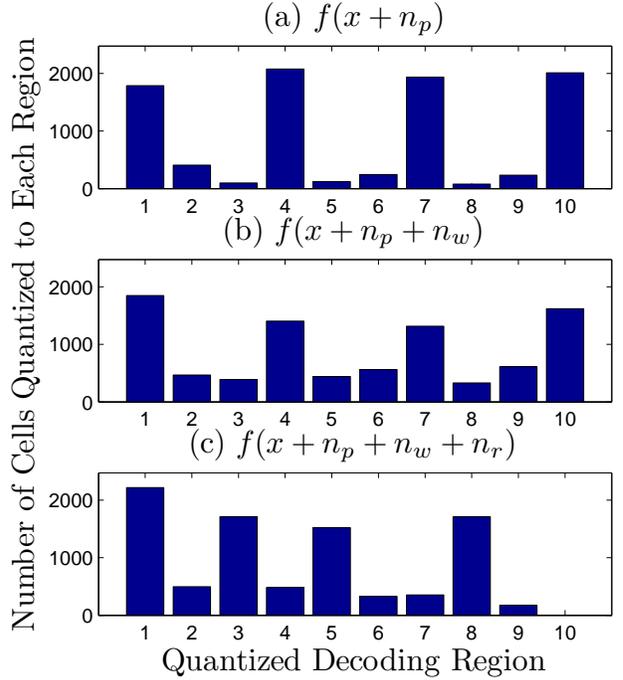}
\caption{Histogram of $9000$ Flash cells with threshold voltage distributions and memory-sensing word-line voltages as shown in Fig. \ref{fig:Noise_PDFs}}
\label{fig:histogram}
\end{figure}

Our dynamic-voltage-level approach determines at regular (though infrequent) intervals how much to increase the voltage levels.  At each channel assessment period, we can numerically solve the general optimization problem of where to place each level to achieve the desired instantaneous capacity with the least growth in $V_{\text{acc}}$.  This requires the ability to assess the wear-out condition of a page.  However, this can be accomplished using the same approach that is used to generate soft information for LDPC decoding \cite{Wang_JASC_2013}.

Estimation of a true distribution from an empirical distribution (or a quantized empirical distribution) is a well known technique.  Fig. \ref{fig:histogram} shows the histogram of threshold voltage regions identified by simulated reading  of $9000$ 4-level MLC Flash memory cells using the word-line voltage model as discussed above and shown in Fig.~\ref{fig:Noise_PDFs}.   The vertical dashed lines in the three threshold voltage distributions shown in Fig.~\ref{fig:Noise_PDFs} are used to produce the histograms in Fig. \ref{fig:histogram}.  These would also be the thresholds used in the limited-precision soft information \cite{Wang_JASC_2013} used by the decoder for two codewords of a rate-$8/9$ LDPC code each protecting $1$ kilobyte of information on a page in memory stored on these $9000$ cells.  Thus, the mechanism to produce this histogram information is already available in Flash memory utilizing multiple reads to obtain soft information for decoding LDPC codes.

Figs.  \ref{fig:histogram} (a) and (b) indicate that the degree of wear-out faced by a page in Flash memory can be estimated from such a histogram with enough accuracy to support the dynamic voltage level approach.  We note that in \cite{LeeTSP2012} the author uses a similar histogram approach to estimate the threshold voltage distribution for the purpose of choosing the best possible quantization thresholds according to \cite{Wang_JASC_2013}.   In \cite{LeeTSP2012}, twelve histogram bins were sufficient to estimate with high precision the means and variances of a mixture of four Gaussians.  Thus, the method of distribution estimation from the histogram is sound.

While information theoretic analysis provides the foundation for this approach, our implementation is practical.  The histogram is generated immediately after a write to avoid confusion with retention loss effects.  Then, whether and how much to increase voltage levels can be determined as a function of how many threshold voltages are outside of the ``correct'' bins.  
This approach can be applied on a block-by-block basis so that constant mutual information can be maintained despite the large variations that have been observed between blocks. 

The histogram approach can also identify and largely mitigate the mean-shift portion of retention loss.  Comparing Figs.  \ref{fig:histogram} (b) and (c) shows the mean shift due to retention loss as a left-shift of the most populated bins, which affects LDPC decoding.  This was also observed in \cite{LeeTSP2012}, where a second cycle of memory sensing follows the histogram-estimated means and variances to obtain optimal limited-precision soft information.  We explore an alternative approach that uses the obtained histogram to adjust how likelihood ratios are assigned to the histogram bins.

The example of Figs. \ref{fig:Noise_PDFs} and \ref{fig:histogram} only considered three quantization thresholds between two adjacent storage levels for simplicity.  Commercial NAND Flash memory chips already support more levels. The Samsung $21$nm $2$ bits/cell chip can support six quantization thresholds between two adjacent storage states~\cite{Kim-Samsung-Flash-JSSC-12}.  Thus we are confident that the mechanism for generating sufficiently rich histograms will exist as a matter of course in future Flash memory systems. 
   
\section{Conclusion}
\label{sec:conclusion}

Using information theory, we have introduced a novel dynamic voltage allocation method  to increase the lifetime of a Flash storage device. A channel model based on voltage-dependent Gaussian noise and Laplace noise is used to demonstrate the numerical results, our idea is applicable to general Flash memory channel model. For the parameters chosen in this paper, the dynamic voltage allocation almost doubles the lifetime of a 4-level Flash memory cell. We expect additional lifetime extension for a general Flash memory channel model. 

In order to obtain the (approximate) mutual information of the Flash memory channel, the noise distribution must be available. We propose estimating the noise distribution by using the quantized soft information obtained during the page-reading process, which is an emerging feature in modern Flash memory devices. 


\bibliographystyle{IEEEtran}
\bibliography{ITA2014}
\end{document}